\input{aipcheck}

\documentclass[
    ,final            
  ]
  {aipproc}

\layoutstyle{6x9}

\begin{document}

\title{The formation of brown dwarfs and low-mass stars by disc fragmentation}



\author{Dimitris Stamatellos}{
  address={School of Physics \& Astronomy, Cardiff University, 5 The Parade, Cardiff CF24 3AA, UK}
}

\author{Anthony P. Whitworth}{
 address={School of Physics \& Astronomy, Cardiff University, 5 The Parade, Cardiff CF24 3AA, UK}
}

\begin{abstract}

We suggest that a high proportion of brown dwarfs are formed by gravitational fragmentation of massive extended discs around Sun-like stars. We argue that such discs should arise frequently, but should be observed infrequently, precisely because they fragment rapidly. By performing an ensemble of radiation-hydrodynamic simulations, we show that such discs typically fragment within a few thousand years to produce mainly brown dwarfs (including planetary-mass brown dwarfs) and  low-mass hydrogen-burning stars. Subsequently most of the brown dwarfs are ejected by mutual interactions. We analyse the properties of these objects that form by disc fragmentation, and compare them with observations.

\end{abstract}

\maketitle


\section{Introduction}

Gravitational fragmentation of massive, extended discs  provides a way to form a large number of brown dwarfs per Sun-like star (Stamatellos et al. 2007). We examine the consequences of this mechanism on the observed properties of brown dwarfs and compare them with observations. More specifically we perform an ensemble of 13 simulations with different numerical realisations of the same star and disc initial conditions, and for the objects produced we compute (i) their mass spectrum, (ii) their radial distribution from the primary star, (iii) their distribution of the orbital planes, (iv) the distribution of the masses and sizes of the brown dwarf discs, and (v) their binary properties.

\section{The simulations}

We consider a 0.7 M$_\odot$ star with a 0.7 M$_\odot$ disc that extends out to 400 AU. The disc surface density drops as $R^{-7/4}$ and the temperature drops as $R^{-0.5}$, where $R$ is the distance from the star on the disc midplane (Stamatellos \& Whitworth, 2008b). 

The simulations are performed in two phases. (i) Until $\sim 80$\% of the disc has been accreted onto the stars (this happens within 20,000 yr) the evolution of the system is followed using SPH, with a new method to capture the thermal and radiative effects when protostellar gas fragments (Stamatellos et al. 2007). The method uses a detailed equation of state for a gas mixture of hydrogen and helium.The effects of the rotational and vibrational degrees of freedom of H$_2$, of H$_2$ dissociation, of H ionisation, and  the first and second ionisation of He are included. Moreover  the opacity changes due to e.g. ice mantle melting, the sublimation of dust, molecular and H$^{-}$ contributions, are also taken into account. The method also includes irradiation by the central star (see Stamatellos \& Whitworth 2008a,b).
(ii) Afterwards the evolution of the system is followed up to 200,000 yr using an N-BODY code.

\begin{figure}
  \includegraphics[height=.22\textheight]{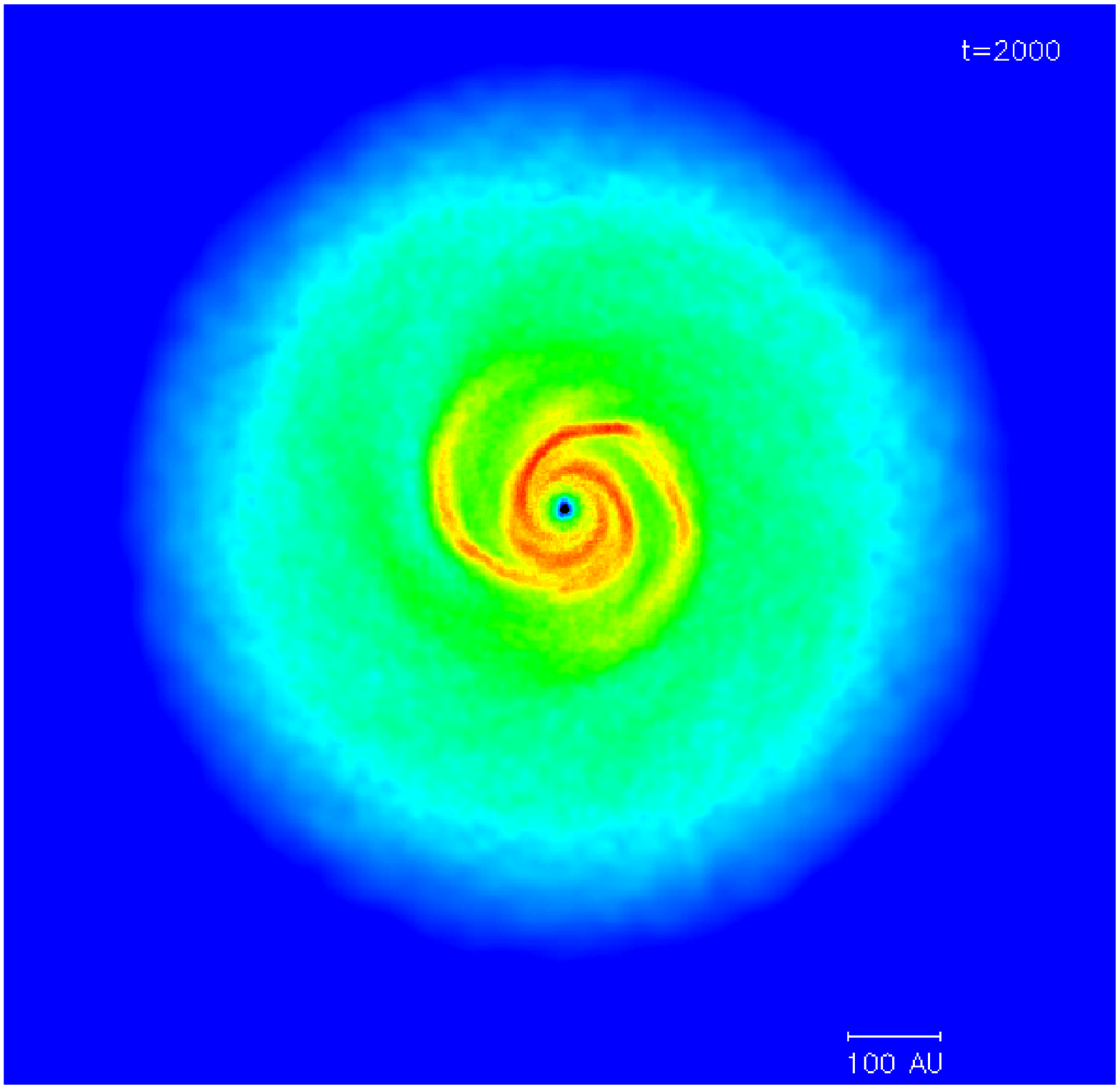}
   \includegraphics[height=.22\textheight]{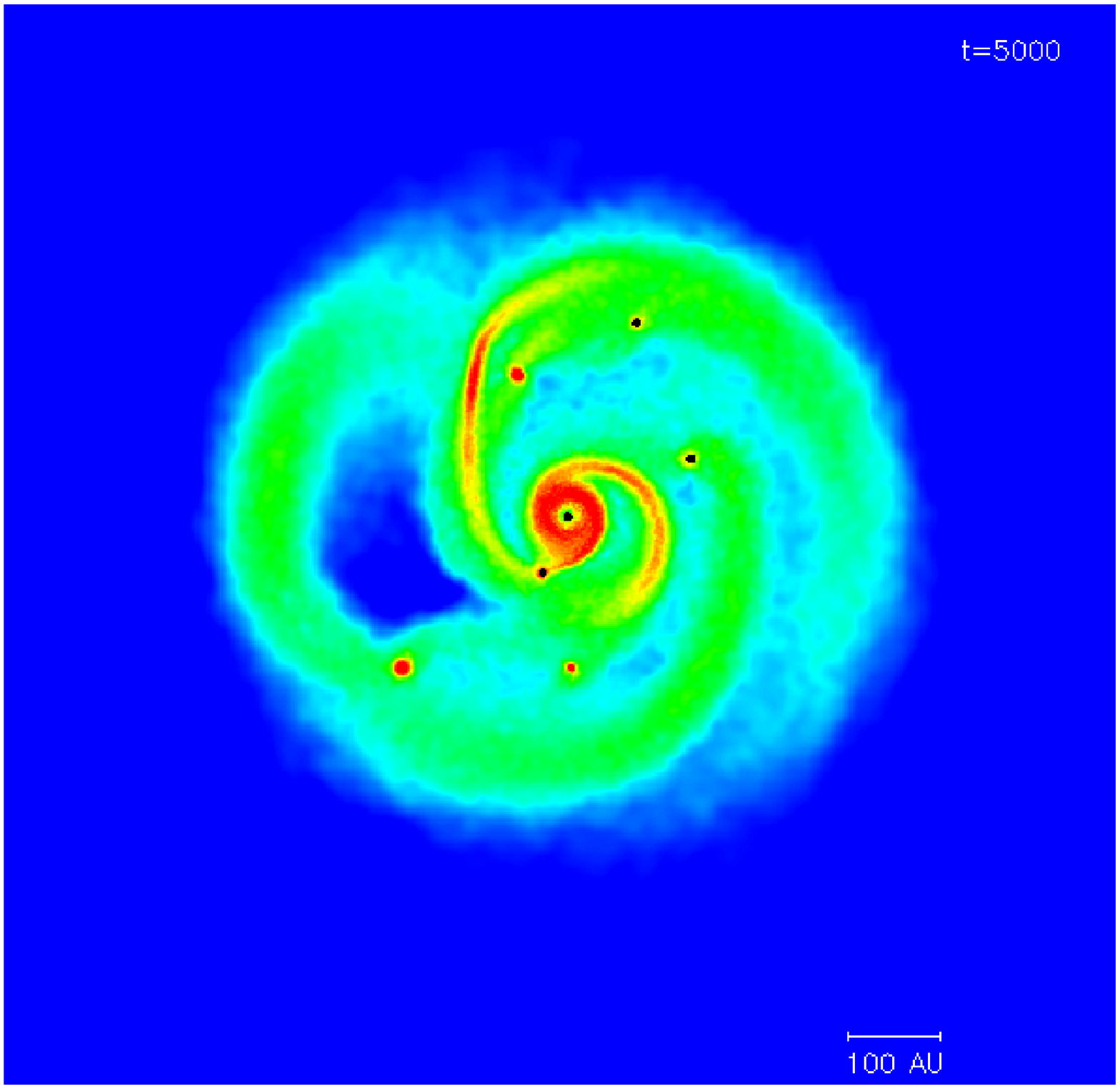}
    \includegraphics[height=.22\textheight]{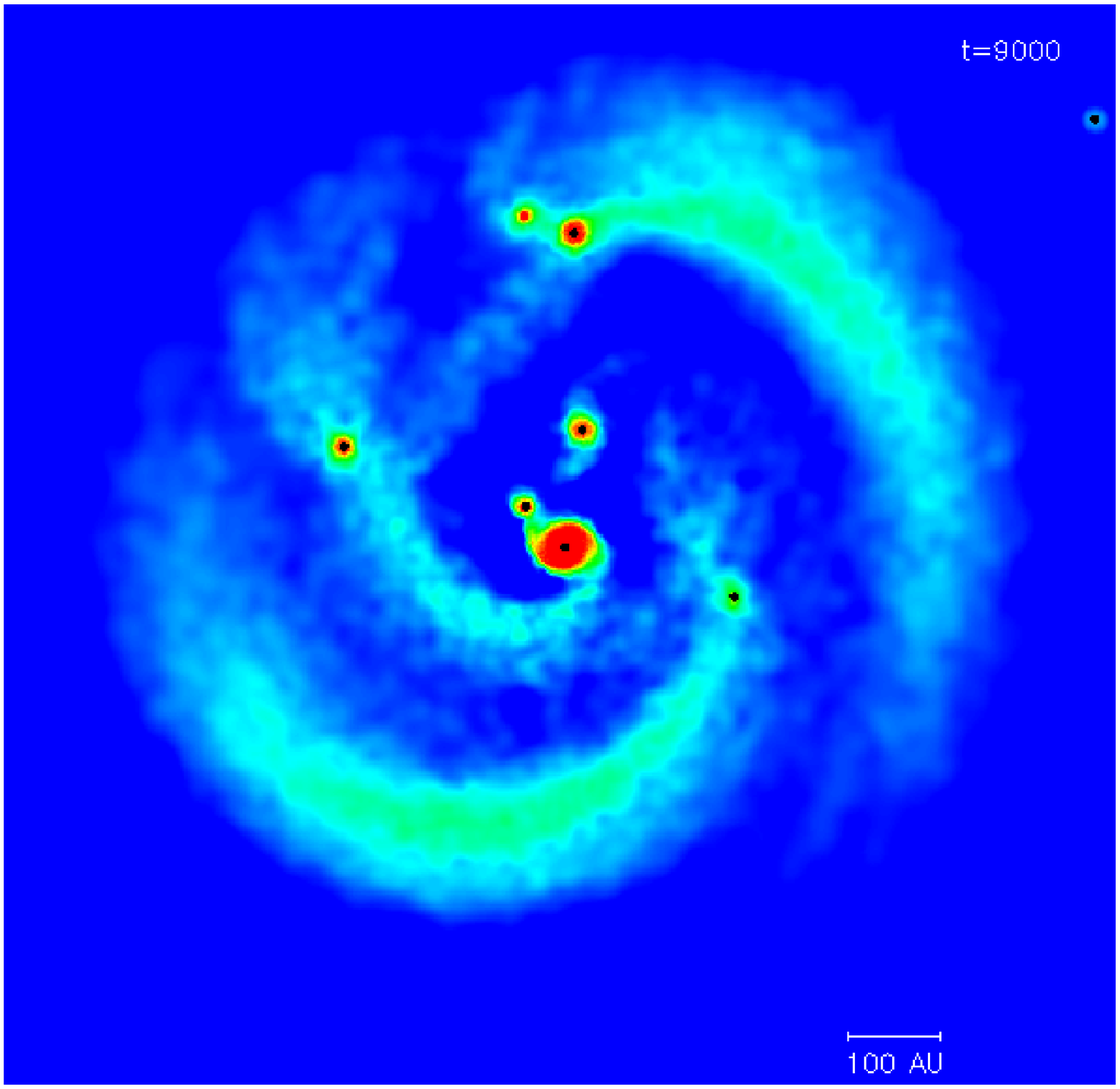}
  \caption{Radiative hydrodynamic simulation of the evolution of  a 0.7-M$_\odot$ disc around a 0.7-M$_\odot$ star. Snapshots of the logarithm of the column density are presented for 3 instances (as marked on each graph). This particular simulation produces 4 brown dwarfs, and 3 low-mass hydrogen burning stars. After 200,000 years, 4 of these objects have been ejected into the field.}
\label{fig:sim}
\end{figure}

The discs quickly (within a few thousand years) fragment (see Fig.~\ref{fig:sim}) forming gravitationally bound objects outside 50 AU, where the conditions for fragmentation are right (i.e. the disc is Toomre unstable {\it and} can cool fast enough) in accordance with our previous simulations and analytic studies (Whitworth \& Stamatellos 2006; Stamatellos et al. 2007; Stamatellos et al. 2008a,b). A total of 96 objects are produced in all of our simulations. More than half of them are ejected into the field. A typical outcome is a Sun-like star with a close low-mass star companion, a wide low-mass star companion, and a wide brown dwarf or brown dwarf binary companion.

\section{Properties of brown dwarfs and low-mass stars  formed by disc fragmentation}

\paragraph{Mass distribution}

Most of the objects produced by disc fragmentation (75\%) are brown dwarf stars (i.e. with masses below the H-burning limit), and 4\% of them are planetary-mass brown dwarfs (i.e. have mass below the D-burning limit). The rest 25\% of the objects produced are low-mass hydrogen-burning stars. More than half (55\%) of the objects produced by disc fragmentation are ejected into the field.

\paragraph{Radial distribution and the brown dwarf desert}

\begin{figure}
  \includegraphics[height=.23\textheight,angle=-90]{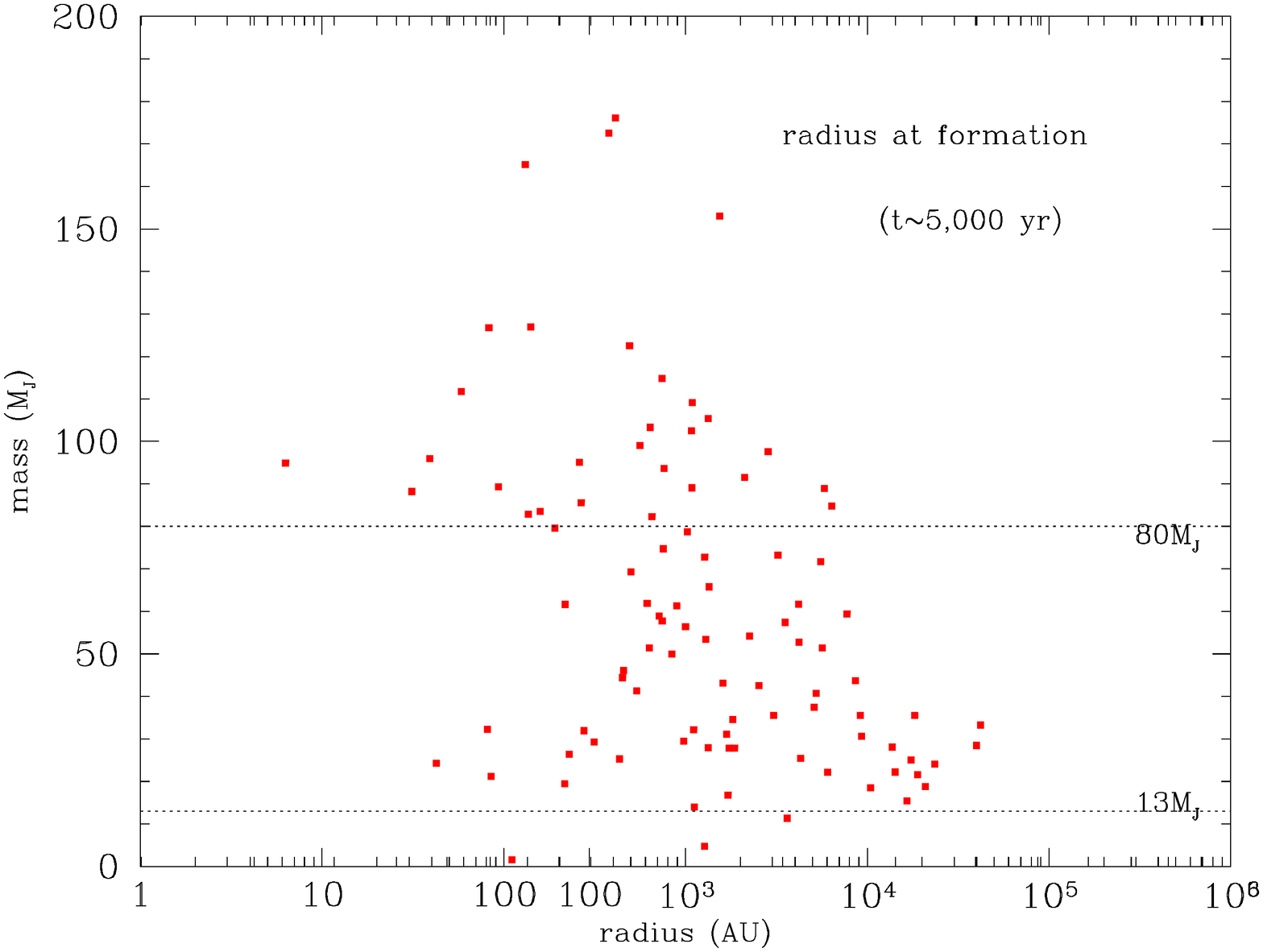}
   \includegraphics[height=.23\textheight,angle=-90]{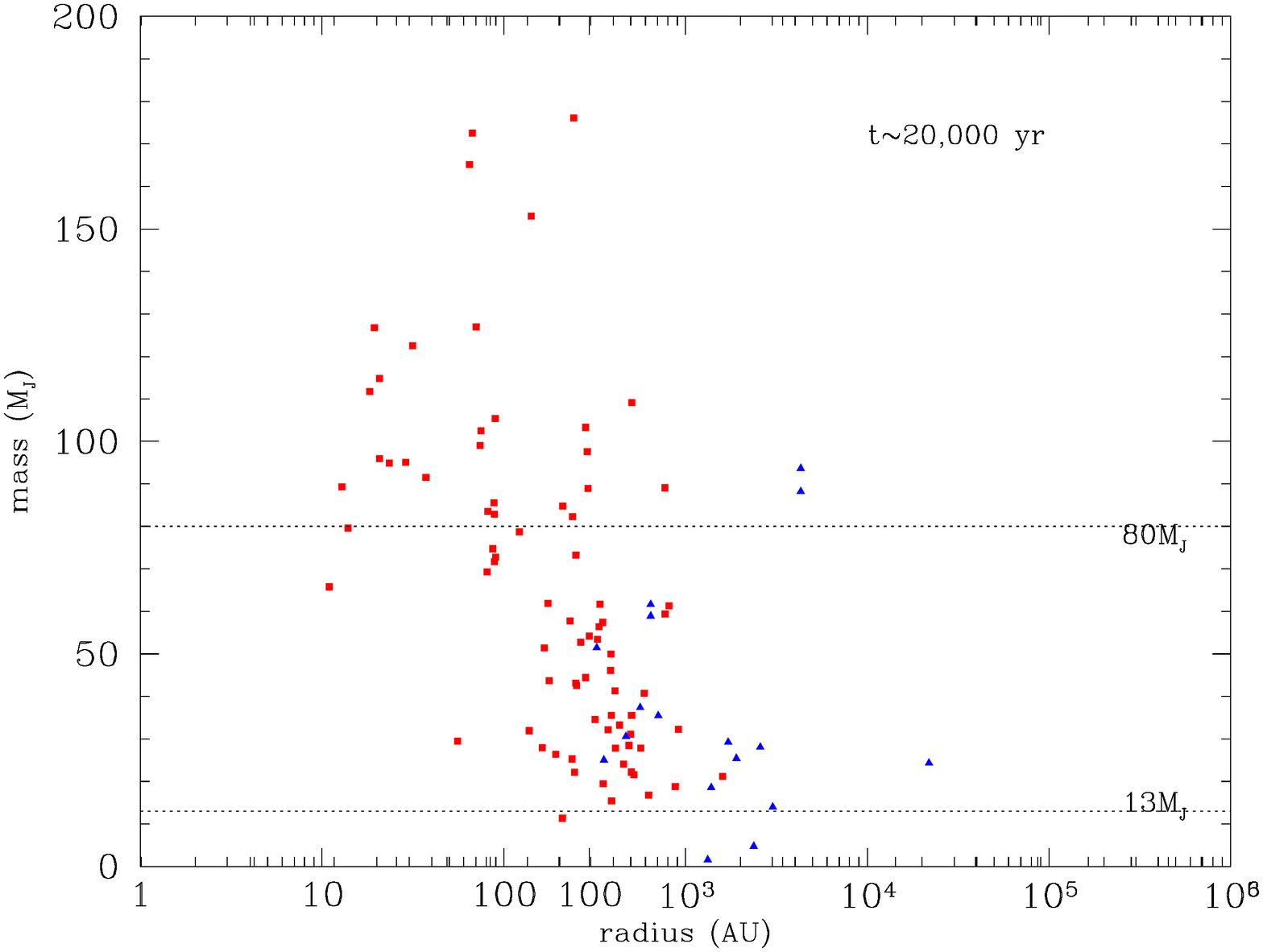}
    \includegraphics[height=.23\textheight,angle=-90]{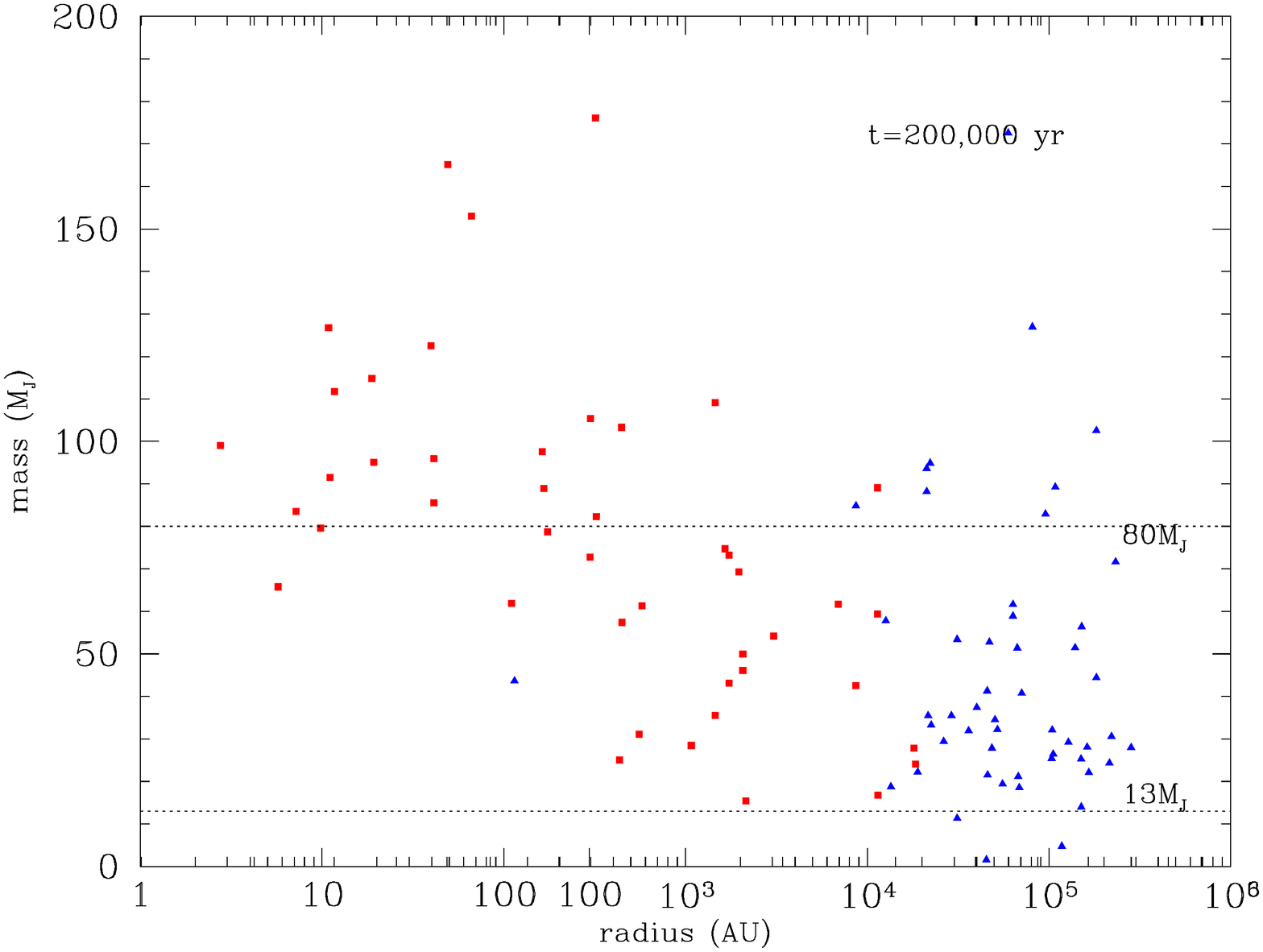}
  \caption{The brown dwarf desert. The formation radius (left), the radius at $t\sim 20,000$ yr (middle), and the the radius at $t=200,000$ yr (right), are plotted against the final object mass.  The blue triangles correspond to objects that are ultimately not bound to the primary star.}
  \label{fig:radius}
\end{figure}

The inner disc cannot cool fast enough  so fragmentation happens only outside 50 AU from the primary star. Objects that form closer to the primary star are generally the most massive ones formed in the disc as (i) they form first, and (ii) they form in a region where there is more material to accrete.  Hence, they become H-burning stars. Brown dwarfs form farther out in the disc. Due to dynamical interactions the low-mass H-burning stars move further in populating the region close to the primary star whereas the brown dwarfs move outwards (see Fig.~\ref{fig:radius}). Hence the mechanism of disc fragmenation provides a natural explanation for the brown dwarf desert (Marcy \& Butler 2000). Our model predicts  that the frequency of brown dwarf companions to Sun-like stars is  8\% within 100 AU from the star and 16\% within 400 AU from the star.

\paragraph{Orbital planes of companions to Sun-like stars}

Initially the orbits of the objects formed in the disc are almost coplanar with the disc of the primary star. Their orbital planes lie within $5^{\rm o}$ from the disc mid-plane. However due to 3 body interactions, in 3-dimensions, after 20,000 yr only 70\% of the orbital planes lie within $5^{\rm o}$ of the disc midplane, and by 200,000 yr this has dropped to 22\%. Hence coplanarity, or the absence of it, cannot confirm or rule out formation by disc fragmentation.

\paragraph{Properties of brown dwarf discs}

Most  brown dwarfs form with discs. These discs have masses of a few M$_{\rm J}$ and radii of a few tens of AU.  These disc masses and radii are consistent with observations, which show discs to be very common around BDs (Klein et al. 2003; Luhman 2004; Luhman et al. 2005; Scholz et al. 2006; Guieu et al. 2007; Riaz \& Gizis 2007, 2008).
Most of the ejected brown dwarfs do not retain their discs. We predict that brown dwarfs that are companions to Sun-like stars are more likely to have discs than brown dwarfs in the field.

\paragraph{Binary properties}

 13 low-mass binaries are formed in our simulations. 27\% of the objects that  form are in binaries corresponding to a low-mass binary fraction of 16\%.  This is comparable with the low-mass binary fraction in Taurus-Auriga ($\stackrel{<}{_\sim} 20\%$, Kraus et al. 2006), Chamaeleon I ($11^{+9}_{-6}\%$, Ahmic et al. 2007), and the field (e.g. $15\pm5 \%$, Gizis et al. 2003). These binaries include star-star (4/13), star-brown dwarf (4/13), brown dwarf -brown dwarf (4/13) and brown dwarf - planetary mass object binaries (1/13).  4 of these binaries are ejected into the field, including both close and wide binaries. We predict that for the binaries remaining bound to the central star, the total mass of the low-mass binary tends to decrease as the distance from the central star increases. Most of the low-mass binaries (55\%) have components with similar masses (q>0.7). Our model suggests that brown dwarfs that are companions to Sun-like stars are more likely to be in brown dwarf- brown dwarf binaries (binary frequency 25\%) than brown dwarfs in the field (binary frequency 8\%).  Burgasser et al. (2005) report a binary fraction of $45^{+15}_{-13}\%$ for BD companions to Sun-like stars, and a binary fraction of only $18^{+7}_{-4}\%$ for BDs in the field.

\paragraph{Free-floating planetary-mass objects}

3 planetary-mass brown dwarfs form in our simulations, and all three of them are ejected into the field. Their masses remain below the D-burning limit because they are ejected from the disc very soon after formation. According to our simulations planetary-mass brown dwarfs in the field should be outnumbered by D-burning brown dwarfs by a factor of $\sim 10$.

\paragraph{Caveats}
 The simulations presented here correspond only to one set of disc and star parameters (i.e. disc density and temperature profiles, star mass, disc mass and disc radius). Hence, the results need to be verified/updated for a wider set of initial conditions. They also do not account for the disc formation around the Sun-like star and disc growth due to accretion of material from the infalling envelope (see Attwood et al. 2008).

\section{Conclusions}
Disc fragmentation is a robust mechanism for the formation of brown dwarfs, as well as planetary-mass objects and low-mass hydrogen-burning stars. It explains successfully properties that are not satisfactorily explained by other formation mechanisms, for example the brown dwarf desert and the statistics of low-mass binary systems. We estimate that even if only a fraction of Sun-like stars host the required massive extended discs needed for the fragmentation scenario to work, this mechanism can produce all the free floating planetary-mass objects, most of the brown dwarfs, and a significant proportion of low-mass hydrogen burning stars.

\bibliographystyle{aipproc}   
\bibliographystyle{aipprocl} 

\def\aap{A\&A}%
\def\mnras{MNRAS}%
\def\aj{AJ}%
\def\actaa{Acta Astron.}%
\def\araa{ARA\&A}%
\def\apj{ApJ}%
\def\apjl{ApJ}%
\def\apjs{ApJS}%
          \def\pasp{PASP}%

\IfFileExists{\jobname.bbl}{}
 {\typeout{}
  \typeout{******************************************}
  \typeout{** Please run "bibtex \jobname" to optain}
  \typeout{** the bibliography and then re-run LaTeX}
  \typeout{** twice to fix the references!}
  \typeout{******************************************}
  \typeout{}
 }

\end{document}